\documentclass[aps,prb,twocolumn,showpacs]{revtex4}
\usepackage{graphicx}
\usepackage{natbib}

\begin{document}

\title{A complete identification of lithium sites in a model of LiPO$_3$
glass: effects of the local structure and energy landscape on ionic
jump dynamics}

\author{Michael Vogel}
\email{mivogel@uni-muenster.de} \affiliation{Department of Chemical
Engineering and Department of Materials Science and Engineering,
University of Michigan, 2300 Hayward, Ann Arbor, MI, 48109,
USA}\altaffiliation{Institut f\"ur Physikalische Chemie,
Westf\"alische Wilhelms-Universit\"at M\"unster, Corensstr.\ 30,
48149 M\"unster, Germany}

\date{\today}

\begin{abstract}
We perform molecular dynamics simulations to study lithium dynamics
in a model of LiPO$_3$ glass at temperatures below the glass
transition. A straightforward analysis of the ionic trajectories
shows that lithium diffusion results from jumps between sites that
are basically unmodified on the time scale of the lithium ionic
relaxation. This allows us a detailed identification and
characterization of the sites. The results indicate that the number
of lithium sites is only slightly bigger than the number of lithium
ions so that the fraction of vacant sites is very limited at every
instant. Mapping the ionic trajectories onto series of jumps between
the sites provides direct access to lithium jump dynamics. For each
site, we determine the mean residence time $\tau_s$ of an ion and the
probability $p_{s}^b$ that a jump from this site to another site is
followed by a direct backjump. While a broad distribution $G(\lg
\tau_s)$ shows that different sites feature very diverse lithium
dynamics, high values of $p_{s}^b$ give direct evidence for
correlated back-and-forth jumps. A strong decrease of $p_{s}^b$ with
increasing $\tau_s$ indicates that the backjump probability depends
on the dynamical state of an ion. Specifically, we find that
correlated back-and-forth jumps are important at short times in the
relaxation process, but not on the time scale of the lithium
relaxation, where the hopping motion resembles a random walk. We
further study how the local glass structure and the local energy
landscape affect lithium jump dynamics. We observe substantial
effects due to the energy landscape, which are difficult to capture
within single-particle approaches. Our results rather imply that
lithium migration is governed by the competition of the ions for a
small fraction of vacant sites in a disordered energy landscape.
Consistently, a statistical analysis shows that a vacancy mechanism
dominates the repopulation of the lithium sites.

\end{abstract}

 \maketitle

\section{Introduction}

The mechanism for fast ion transport in glasses has challenged the
combined efforts of industrial and scientific research over the last
decades.~\cite{BIM,Ngai,BFI} It is now well established that the
macroscopic charge transport results from hopping of mobile ions
between localized states in the glassy matrix.~\cite{BIM,BFI}
However, the mechanism of this hopping motion is complex and several
coexisting models seek to rationalize the very rich phenomenology of
ionic glasses. While several authors postulated that ionic migration
is governed by the Coulomb interactions among the mobile
ions,~\cite{Funke1,Elliott,Funke2} others stressed the relevance of
the disordered energy landscape provided by the glassy
matrix,~\cite{Stevels,Ingram,Dyre1} and percolation approaches were
applied for a quantitative description of ionic
diffusion.~\cite{Svare,Sen,Baranovskii} The combined effect of these
interactions was also regarded as
important.~\cite{Maass1,Maass2,Greaves1,Dietrich} In addition, it was
argued that structural properties, e.g., microsegregation of the
mobile ions, play a major role for ion
dynamics.~\cite{Greaves1,Greaves2} Also, the relevance of substantial
relaxation of the glass structure below the glass transition
temperature $T_g$ was emphasized.~\cite{Maass3}

To decide between different models of ion transport in glasses,
detailed information about the ionic motion is required. For example,
it is interesting whether all ions make the same contribution to the
macroscopic charge transport or whether a distribution of mobilities
exists. Further, the relevance of correlated back-and-forth jumps
needs to be quantified. Due to this effect, which is one of the
cornerstones of various modelling approaches, the dc conductivity
$\sigma_{dc}$ can be much smaller than expected based on the ionic
jump rates. These points are related to the question about the origin
of the nonexponential ionic relaxation, observed in electrical and
mechanical relaxation studies.~\cite{Ngai,Moynihan,Liu,Green1,Green2}
Such behavior can result from two fundamentally different
scenarios.~\cite{12} In the heterogeneous scenario, all particles are
random walkers, but a distribution of jump rates exists. In the
homogeneous scenario, all particles obey the same relaxation function
that is intrinsically nonexponential due to correlated back-and-forth
jumps. Additionally, it is intriguing to study how ion dynamics
depends on the local glass structure.

Measuring the frequency dependent electric conductivity
$\sigma(\nu)$, an experimental study of ion dynamics is possible.
While the frequency independent value at low frequencies,
$\sigma_{dc}$, is an important material parameter, a strong
dispersion at higher frequencies indicates the presence of correlated
back-and-forth motions.~\cite{Funke2,Jonscher,Martin,Funke3,Roling}
Multidimensional nuclear magnetic resonance (NMR) experiments provide
direct access to ionic jump
dynamics.~\cite{Bohmer1,Bohmer2,Vogel_JNCS,Vogel_PCCP,Vogel_PRB}
Applications on crystalline and glassy silver ion conductors showed
that the nonexponential depopulation of the silver sites is due to a
distribution of jump rates, i.e., dynamic heterogeneity, rather than
to an intrinsic noneponentiality.~\cite{Vogel_PCCP,Vogel_PRB} In
other words, no evidence for back-and-forth jumps was found, at
variance with conclusions from electric conductivity studies.

In molecular dynamics (MD) simulations, the trajectories of all
particles are known and, thus, unique microscopic insights are
available. Applications on ionic glasses confirmed that the dynamics
of the mobile ions can be decomposed into vibrations about sites and
occasional jumps between
them.~\cite{Smith,Bala,Karth,Timpel,Kieffer,Park,Horbach1,Habasaki,Cormack,Heuer1,Heuer2,Vogel}
Most simulation studies of ionic jump dynamics focused on alkali
silicate glasses. For lithium silicate glasses, it was demonstrated
that both homogeneous and heterogeneous dynamics contribute to the
nonexponentiality of ionic relaxation.~\cite{Heuer2,Habasaki} For
sodium silicate glasses, no evidence for back-and-forth jumps was
found, but there are dynamic
heterogeneities.~\cite{Jund,Sunyer1,Kob,Horbach2,Sunyer2}
Specifically, fast ion transport was observed along preferential
pathways, where, however, the presence of these channels was not
attributed to a microsegregation of the sodium ions, as proposed by
other workers.~\cite{Greaves1,Vessal}

Very recently, we performed MD simulations to study ion dynamics in
an alkali phosphate glass.~\cite{Vogel} Based on multitime
correlation functions, we demonstrated that the nonexponential
lithium relaxation in LiPO$_3$ glass results from both correlated
back-and-forth jumps and a broad distribution of jump rates. A
quantitative analysis showed that the relevance of the heterogenous
contribution increases with decreasing $T$. In addition, we observed
an exchange between fast and slow lithium ions that occurs on the
timescale of the jumps themselves. Thus, the dynamic heterogeneities
are short lived, indicating that sites featuring fast and slow
lithium dynamics, respectively, are intimately mixed.

Heuer and coworkers~\cite{Lammert} used a computational approach to
demonstrate that determination of the lithium sites in lithium
silicate glasses provides new insights into the mechanism for ionic
diffusion. A complete identification of the sites from the ionic
trajectories showed that the number of sites is only slightly bigger
than the number of ions, implying that ion dynamics is most
appropriately described in terms of mobile vacancies. For a detailed
study of the jump diffusion mechanism, the alkali trajectories were
mapped onto sequences of jumps between the sites.~\cite{Lammert} An
analysis of these sequences revealed that the tendency for
back-and-forth jumps depends on the dynamical state of an ion.
Specifically, the back-jump probabilities are high when the lithium
ions have escaped from sites characterized by comparatively short
residence times.

In the present MD simulation approach, we investigate lithium jump
dynamics in LiPO$_3$ glass via identification of the lithium sites.
In particular, we study the effects of the local glass structure and
the local energy landscape on the lithium hopping motion. To unravel
the origin of the pronounced dynamic heterogeneities observed for
this model glass in our previous work,~\cite{Vogel} we analyze which
factors determine the residence time at a site. Such analysis is also
performed for the correlated back-and-forth jumps so that the present
study may shed a light on the question why previous experimental and
computational studies came to different conclusions about the
relevance of this phenomenon. Furthermore, we use a statistical
approach to ascertain the dominant mechanism for the repopulation of
the lithium sites. All analyses are carried out in a temperature
range, where the phosphate glass matrix, apart from local
fluctuations, is rigid on the $20\,\mathrm{ns}$-time scale of our
simulation, while the lithium ionic subsystem can still be
equilibrated.

\section{Model and simulation details}

The interactions of the ions in the studied model of LiPO$_3$ glass
are described by the potential
\begin{equation}\label{potential}
\Phi_{\alpha\beta}(r)=\frac{q_{\alpha}q_{\beta}\,e^2}{r}+
A_{\alpha\beta}\,\exp\left(-r/\rho\right),
\end{equation}
where $e$ is the elementary charge and $r$ denotes the distance
between two ions of type $\alpha$ and $\beta$, respectively
($\alpha,\beta\!\in\! \{\mathrm{Li,O,P}\}$). Karthikeyan et
al.\cite{Karthikeyan} adjusted the parameters of this potential to
enable a realistic description of the structure of LiPO$_3$ glass.
They showed that, though there may be some deviations in the
intermediate range order,\cite{Alam} the simulated glass consists of
well defined phosphate tetrahedra that are connected by two of their
corners to form long chains and/or rings,~\cite{Karthikeyan} in
agreement with experimental findings.~\cite{Hoppe,Brow,Wullen} In
previous work,~\cite{Vogel} we studied the dynamical behavior of this
model using a slightly modified set of potential parameters. We found
that the dynamics of the atomic species decouple with decreasing $T$.
While the structural relaxation of the phosphate network freezes in
at a computer glass transition temperature
$T_g\!\approx\!1000\mathrm{\,K}$, fast lithium ionic diffusion takes
place at $T\!<\!T_g$. Thus, it is possible to equilibrate the lithium
ionic subsystem at $T\!<\!T_g$ and to study ionic migration in a
glassy matrix. Detailed analysis of the lithium dynamics showed that
the dynamical behavior of the model resembles that of LiPO$_3$
glass.~\cite{Vogel}

In the present work, we identify the lithium sites in this model of
LiPO$_3$ glass. For this purpose, we extend the temperature range of
our previous study~\cite{Vogel} to lower $T$, where lithium dynamics
is well described as hopping motion. We perform MD simulations in the
NVE ensemble with $\rho\!=\!2.15\mathrm{\,g/cm^3}$ and $N\!=\!800$ so
that the number of lithium ions $N_{\mathrm{Li}}\!=\!160$. Such
moderate system size allows us to equilibrate the lithium ionic
subsystem at sufficiently low $T$, while major finite size effects
are absent.~\cite{Vogel} The density is chosen based on the
experimental value at room temperature,
$\rho\!=\!2.25\,\mathrm{g/cm^3}$,~\cite{Muru} together with the
thermal expansion coefficient of LiPO$_3$ glass.~\cite{English} The
equations of motion are integrated using the velocity Verlet
algorithm with a time step of $2\mathrm{\,fs}$, periodic boundary
conditions are applied and the Coulombic forces are calculated via
Ewald summation.~\cite{Vogel}

Our simulations start from a configuration that was obtained
previously~\cite{Vogel} by equilibrating the phosphate network
slightly above $T_g$, quenching into the glass and equilibrating the
lithium ionic subsystem at $T\!=\!590\mathrm{\,K}$. To set the target
temperatures of the present study, $T\!=\!390\mathrm{\,K}$,
$T\!=\!450\mathrm{\,K}$ and $T\!=\!520\mathrm{\,K}$, we first
propagate this configuration while periodically rescaling the
velocities. Afterwards, equilibration periods of lengths
$t_{eq}\!=\!30\!-\!40\mathrm{\,ns}$ without velocity scaling are
applied at each $T$. Finally, we perform production runs of length
$t_{sim}\!=\!20\mathrm{ns}$. For $T\!=\!520\mathrm{\,K}$, all lithium
ions exit their sites during the equilibration period $t_{eq}$,
indicating that the lithium ionic subsystem is well equilibrated. For
the lower $T$, this condition cannot be met. However, apart from
fluctuations, the temperature is constant after $t_{eq}$ and, hence,
the population of the lithium sites has reached its equilibrium
distribution. Such behavior becomes reasonable based on our result
that the number of lithium sites is very limited, see below, so that
sites with long residence times and low energies are occupied in any
case and equilibration only requires the repopulation of sites with
short residence times and high energies, between which numerous
exchange processes take place during the applied equilibration
periods.

To further check the equilibration of the lithium ionic subsystem, we
calculate the incoherent intermediate scattering functions of the
lithium ions, see Fig.~\ref{fig3}, and determine the mean time
constants of the decays, as was done previously.~\cite{Vogel}
Comparison of previous and present results shows that all data at
$T\!<\!T_g$ are well described by an Arrhenius law with activation
energy $E_a\!=\!0.47\mathrm{\,eV}$. Fitting data for both $T\!>\!T_g$
and $T\!<\!T_g$, we found a bigger value $E_a\!=\!0.62\mathrm{\,eV}$
in our previous study.~\cite{Vogel} Detailed inspection, however,
shows that this difference can be traced back to a change in the
temperature dependence at
$T\!\approx\!T_g\!\approx\!1000\mathrm{\,K}$. Based on these
findings, we conclude that the lithium ionic subsystem is
equilibrated at all $T$ studied here.

\section{Identification of lithium sites}

Following Heuer and coworkers,~\cite{Lammert} a straightforward
algorithm is applied to identify the lithium sites based on the MD
trajectories. In detail, we first divide the simulation box into
cubic cells of size $(0.3\mathrm{\,\AA})^3$, which is sufficiently
small so as to resolve the shape of the sites. For each $T$, we then
determine the number of times, $m$, the cells are visited by a
lithium ion during $t_{sim}\!=\!20\mathrm{ns}$, where configurations
separated by a time increment $\Delta t\!=\!0.2\mathrm{\,ps}$ are
analyzed. The visited cells thus constitute the lithium sites and the
pathways between them. To eliminate the pathways, only cells with
$m\!>\!m_0$ are considered in a following cluster analysis, where
cells sharing a common face are grouped into one site. The value of
$m_0$ is determined by the criterion that the number of distinct
lithium sites be maximum. While lithium sites that are occupied less
frequently are not detected for high values of $m_0$, different sites
merge for small values of $m_0$ due to inclusion of the connecting
pathways.

Having identified the lithium sites, the migration of the lithium
ions can be mapped onto sequences of jumps between these sites. When
an ion has left a site, it can either move to a new site or return to
the old one. While we record the former event as a jump, we dismiss
the latter and regard the residence at the particular site as not
interrupted. In this way, we can exclude the effect of occasional
large amplitude vibrations, which lead to exploration of volume
outside the sites, but not to long-range transport. Thus, the
residence time of a lithium ion at a site is defined as the interval
between the time when the ion jumps into this site and the time when
the ion exits this site in order to enter \emph{another} site. Based
on the residence times, we calculate both the mean residence time
$\tau_s$ at a site and the mean number of lithium ions at a site, or,
equivalently, the occupation number $n_s$. To determine $\tau_s$, a
sufficiently large number of jump processes has to be analyzed. This
is possible for $T\!=\!520\mathrm{\,K}$, whereas some sites are
repopulated just a few times during $t_{sim}$ at the lower $T$. To
reduce the effect of the limited time window, we multiply the
residence times at the beginning and at the end of the production run
by a factor of two when calculating $\tau_s$.

\begin{figure}
\includegraphics[angle=0,width=8.5cm,clip]{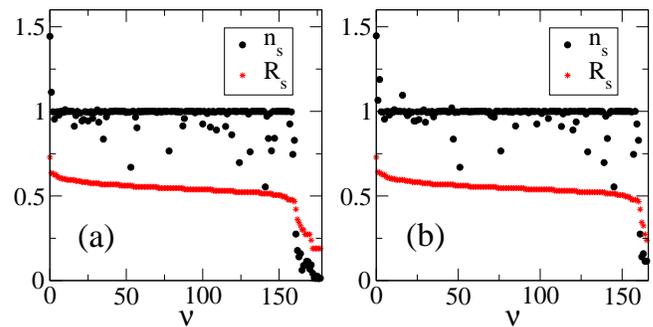}
\caption{Effective radius $R_s\!=\!(3V/4\pi)^{1/3}$ of the lithium
sites and mean number $n_s$ of lithium ions at a site. (a) Results
for 178 sites obtained from a preliminary analysis at
$T\!=\!390\mathrm{\,K}$. (b) Results for the final set of 166 sites
determined by eliminating satellite and transition sites, see text
for details. In both panels, the sites are sorted according to their
effective radius, which is plotted in units of $\mathrm{\AA}$, and
$\nu$ denotes the site index.}\label{fig1}
\end{figure}

Heuer and coworkers~\cite{Lammert} observed that some small sites
resulting from the described algorithm do not serve as independent
sites, but they are satellites of larger sites or saddle-like states
that are visited for a short period of time during the transitions
between larger sites. Therefore, these small sites were excluded from
their analysis. In view of these results, we determine the lithium
sites in two steps. First, we identify the maximum number of sites at
each $T$ and study their nature in a preliminary analysis. Based on
the findings, we then eliminate satellite and transition sites and
perform the final analysis for the resulting set of sites, which
turns out to be temperature independent.

The preliminary analysis yields between 176 and 193 lithium sites at
the studied $T$. While 166 sites are present at all $T$ (A sites),
there is an additional temperature dependent set of sites (B sites).
Specifically, when the temperature is increased, some sites merge and
some new sites become available. Further analysis showed that the B
sites exhibit distinctly smaller occupation numbers $n_s$ and
effective radii $R_s\!=\!(3V/4\pi)^{1/3}$ than the A sites so that
they contribute to the drops observed for these quantities in
Fig.~\ref{fig1}(a). The B sites are also separated by unusually
small center-of mass distances $r_{cm}$ from at least one other site.
To further study the nature of the lithium sites, we calculate the
occupation number of site $i$ under the condition that site $j$ is
occupied, $ n_{ij}$, and normalize it by the regular occupation
number of site $i$, $n_i$. Then, a ratio $n_{ij}/n_i\!\approx\!1$ is
observed for all A sites, indicating an independent occupation of
these sites, while values $n_{ij}/n_i\!\ll\!1$ are found for the B
sites and, hence, their population is suppressed by the population of
other sites. Finally, from inspection of the probabilities $p_{ij}$
that a jump from site $i$ leads to a site $j$, it becomes clear that
many B sites are predominantly exited towards a particular other
site, as is expected for satellite sites.

\begin{figure}
\includegraphics[angle=270,width=8.5cm,clip]{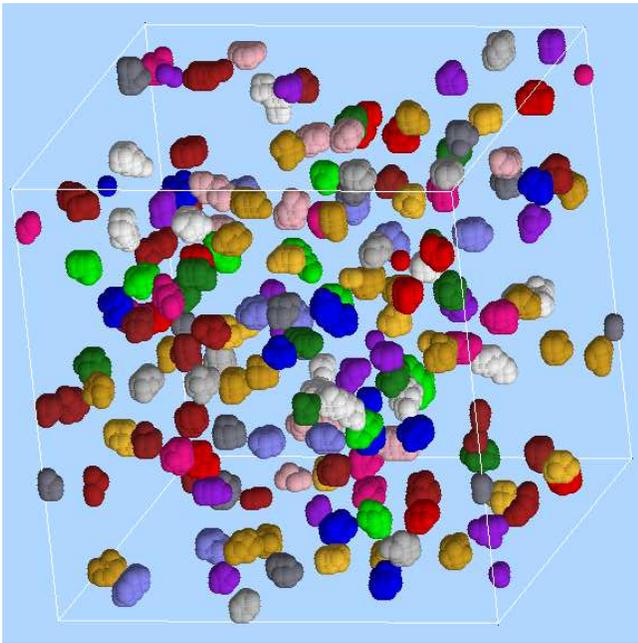}
\caption{Lithium sites in LiPO$_3$ glass. The sites were obtained by
dividing the simulation box into small cubic cells and counting the
number of times a particular cell is visited by a lithium ion in the
course of the simulation. Frequently visited, face sharing cells were
then grouped into one site. Here, the cells are shown as spheres,
where cells constituting distinct lithium sites are given different
colors.}\label{fig2}
\end{figure}

These findings show that some of the lithium sites resulting from our
preliminary analysis are indeed satellite or transition sites and,
hence, their explicit consideration is not necessary.~\cite{Lammert}
To determine the ``relevant'' lithium sites, we start from the set
identified at $T\!=\!390\mathrm{\,K}$ and combine any sites $i$ and
$j$ if at least three of the following four criteria are fulfilled:
(i) the sites merge at higher $T$, (ii) the sites are separated by a
center-of mass distance $r_{cm}\!<\!2.0\mathrm{\,\AA}$, which is
smaller than the minimum Li-Li interatomic distance, i.e.,
$g_{LiLi}(r\!<\!2.0\mathrm{\,\AA})\!=\!0$, (iii) a ratio $n_{ij}/n_i
\!<\!0.3$ indicates that the sites are not independently occupied and
(iv) a value $\mathrm{max}\,[p_{ij},p_{ji}]\!>\!0.5$ shows that jumps
from one of the sites predominantly lead to the other. As a result,
we find that all B sites identified at $T\!=\!390\mathrm{\,K}$ are
associated with an A site, leading to a number of 166 independent
lithium sites. This final set of lithium sites is shown in
Fig.~\ref{fig2}. We see that the sites are compact, mostly globular
objects, which are well separated from each other. All further
analysis is based on these sites. However, we determined that our
conclusions are not affected when the following analysis is performed
on the preliminary sets of sites. Moreover, we ensured that the first
and the second half of our production runs, respectively, yield
nearly identical sets of sites.

\section{Results}

\subsection{Properties of the lithium sites}

Analyzing the ionic trajectories, we find that, for all studied $T$,
more than 96\% of the lithium positions lie within the volume
constituted by the 166 lithium sites, i.e., within a volume fraction
of only 1\%. These percentages show that our approach allows us a
complete identification of the lithium sites. In addition, they imply
that the positions of the lithium sites are temperature and time
independent, which in turn suggests that structural relaxation of the
phosphate matrix plays no substantial role on the time scale of our
simulation. Consistently, we observe that the lithium sites exhibit
temperature independent properties, i.e., sites with relatively long
(short) mean residence times $\tau_s$, high (low) occupation numbers
$n_s$, etc.\ at a given $T$ show the same features at the other $T$ 
studied here.

To investigate the dynamics of the phosphate matrix in more detail,
we calculate the mean square displacements of the oxygen and
phosphorus atoms, $r^2_\mathrm{O}(t)$ and $r^2_\mathrm{P}(t)$,
respectively. Values
$r^2_{\mathrm{O,P}}(10\,\mathrm{ns})\!<\!0.6\mathrm{\,\AA}^2$
indicate that substantial structural relaxation of the phosphate
matrix is absent at the studied $T$. To obtain further insights, we
determine the number of phosphorus atoms in the first neighborshells
of the oxygen atoms, i.e., we identify bridging oxygens (BO) and
nonbridging oxygens (NBO). For all $T$, we find that more than 98\%
of the oxygen atoms have the same number of phosphorus neighbors at
the beginning and at the end of our $20\,\mathrm{ns}$-time window
and, hence, each oxygen atom has a preferred phosphorus coordination.
However, there are temporary matrix fluctuations. For example, when
we analyze configurations at 1000 equidistant times during
$t_{sim}\!=\!20\mathrm{\,ns}$, we observe that the fraction of oxygen
atoms that \emph{never} changes the number of phosphorus neighbors is
reduced to 75\% for $T\!=\!520\mathrm{\,K}$.

Motivated by the finding that the spatial distribution of NBO is
basically static on the time scale of our simulation, we next study
the relative positions of lithium sites and NBO. For this purpose, we
identify the oxygen atoms in the first neighborshells of
lithium ions residing at a site and determine the fraction of NBO
among this subset of oxygens. A statistical analysis shows that
the fraction amounts to more than 96\% at the studied $T$ and, hence,
a lithium ion at a site is predominantly surrounded by NBO. In other
words, the lithium sites are located in regions with high local
concentrations of NBO. For comparison, as expected for the
meta-phosphate composition, the fraction of NBO among all oxygens in
the system amounts to $2/3$.

Figure \ref{fig1}(b) displays the effective radii $R_s$ of the 166
lithium sites together with their occupation numbers $n_s$ at
$T\!=\!390\mathrm{\,K}$. We see that most sites are characterized by
radii $R_s\!\approx\!0.5\mathrm{\AA}$, indicating that the sites are
compact objects of similar size, consistent with the appearance of
Fig.~\ref{fig2}. Further, we observe that the vast majority of the
sites exhibits occupation numbers $n_s\!\lesssim\!1$, i.e., they can
accommodate only one ion at a time, so that the total number of
available sites, $N_s$, is approximately given by the number of
identified sites. Hence, the number of sites, $N_s\!\approx\!166$, is
only slightly bigger than the number of ions,
$N_{\mathrm{Li}}\!=\!160$. Consistently, Heuer and
coworkers~\cite{Lammert} reported $N_s\!\approx\!N_{\mathrm{Li}}$ for
Li$_2$SiO$_3$ glass and proposed to describe lithium dynamics in
terms of mobile vacancies. Moreover, Dyre~\cite{Dyre2} argued based
on general reasoning that a minimum number of alkali sites is formed
during the freezing of the network at $T_g$ due to energetic reasons.

\begin{figure}
\includegraphics[angle=0,width=8.5cm,clip]{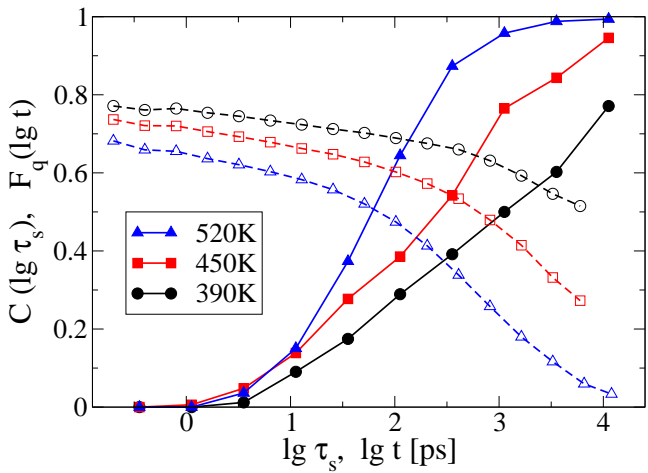}
\caption{Solid symbols: Cumulative distribution $C(\lg \tau_s)$
characterizing the mean residence times of the lithium ions at the
sites. Open symbols: Incoherent intermediate scattering function of
the lithium ions, $F_q(\lg t)$, for
$q\!=\!2\pi/r_{LiLi}\!=\!2.3\mathrm{\,\AA^{-1}}$.}\label{fig3}
\end{figure}

Next, we characterize the mean residence times $\tau_s$ at the
lithium sites by the cumulative distribution $C(\lg \tau_s)$. The
latter is related with the distribution of mean residence times,
$G(\lg \tau_s)$, by
\begin{displaymath}
C(\lg \tau_s)\!=\!\int_{-\infty}^{\lg \tau_s}G(\lg
\tau_s^{\prime})\;d\lg\tau_s^{\prime}
\end{displaymath}
so that the cumulative distribution measures the fraction of sites
with residence times smaller than $\tau_s$. In Fig.~\ref{fig3}, we
see that $C(\lg \tau_s)$ shifts to longer times with decreasing $T$,
reflecting the slowing down of lithium jump dynamics. Furthermore,
the rise of $C(\lg \tau_s)$ is more gradual at lower $T$ and, hence,
the distribution $G(\lg \tau_s)$ broadens upon cooling. In
particular, the increase of $C(\lg \tau_s)$ extends over more than
four orders of magnitude at $T\!=\!390\mathrm{\,K}$, indicating that
the sites feature very diverse lithium jump dynamics, i.e.,
pronounced dynamic heterogeneities exist. In addition, it becomes
clear from Fig.~\ref{fig3} that about 20\% of the lithium ions
uninterruptedly reside at the same site during our production run for
$T\!=\!390\mathrm{\,K}$, and, thus, the mean residence time of these
sites cannot be determined. In the following analysis, we use
$\tau_s\!=\!40\mathrm{\,ns}\!\equiv\!2t_{sim}$, which may somewhat
affect the results for $T\!=\!390\mathrm{\,K}$.

For comparison, we consider the incoherent intermediate scattering
function of the lithium ions,
\begin{displaymath}
F_q(t)=\langle\,\cos\{\,\vec{q}\,[\vec{r}(t_0\!+\!t)\!-\!\vec{r}(t_0)]\}\,\rangle.
\end{displaymath}
Here, $\vec{r}(t)$ is the position of a lithium ion at a time $t$ and
the brackets $\langle\dots\rangle$ denote the ensemble average. To
study lithium dynamics on the length scale of the Li-Li interatomic
distance, $r_{LiLi}\!=\!2.7\mathrm{\AA}$, we use an absolute value of
the wave vector $q\!=\!2\pi/r_{LiLi}$, as was done
previously.~\cite{Vogel} In Fig.~\ref{fig3}, we show the temperature
dependence of $F_q(\lg t)$. Comparing $C(\lg \tau_s)$ and $F_q(\lg
t)$ for $T\!=\!450\mathrm{\,K}$, we see that the latter
function does not decay to zero on a time scale of $10\mathrm{\,ns}$
($\lg t[\mathrm{ps}]\!=\!4$) and, hence, a significant fraction of 
lithium ions still occupies the initial site, whereas $C(\lg
\tau_s[\mathrm{ps}]\!=\!4)\!\approx\!1$ indicates that the residence
times at nearly all sites were interrupted by jumps to other sites
during this time window. In combination, these findings suggest that
there is a substantial fraction of unsuccessful escape processes,
i.e., the ions jump back to the initial site. Thus, the results of
this section confirm our previous findings~\cite{Vogel} that both
dynamic heterogeneities and correlated back-and-forth jumps are
important features of lithium jump dynamics in LiPO$_3$ glass. As will be
demonstrated below, the present approach allows us to study the
origin of these phenomena in detail.

\subsection{Mechanism for the lithium jumps}

To analyze the mechanism for the lithium jumps, we determine the mean
jump time $\tau_j$ and the mean jump length $l_j$, where we define
the former as the time an ion spends outside any site when it moves
from one site to another and the latter as the center-of mass
distance of two consecutively visited sites. At the studied $T$, we
find mean jump times $\tau_j\!=\!0.6\!-\!1.7\mathrm{\,ps}$ that are
much shorter than the mean residence times $\tau_s$, cf.\
Fig.~\ref{fig3}, confirming that jump diffusion occurs. The mean jump
lengths $l_j\!=\!3.2\!-\!3.7\mathrm{\,\AA}$ are only slightly bigger 
than the interatomic distance
$r_{\mathrm{LiLi}}\!=\!2.7\mathrm{\,\AA}$ and, thus, the lithium ions
predominantly jump to neighboring sites.

When we consider the very limited number of lithium sites
$N_s\!\approx\!N_{\mathrm{Li}}$, there are two possible mechanisms
for the repopulation of the sites. On the one hand, one can imagine a
vacancy mechanism where a site is first exited by ion A and
immediately afterwards occupied by ion B. Indeed, studying examples
of ionic trajectories, a vacancy-like mechanism was observed for a
silicate glass.~\cite{Cormack} On the other hand, it is possible that
ion B enters an occupied site and ``kicks out'' ion A. Both scenarios
can be distinguished when we determine the probability that a lithium
jump leads to an empty site. We find that this probability is higher
than 84\% in the considered temperature range, indicating that a
vacancy mechanism dominates the repopulation of the lithium sites.

\subsection{Origin of the dynamic heterogeneities}

\begin{figure}
\includegraphics[angle=0,width=8.5cm,clip]{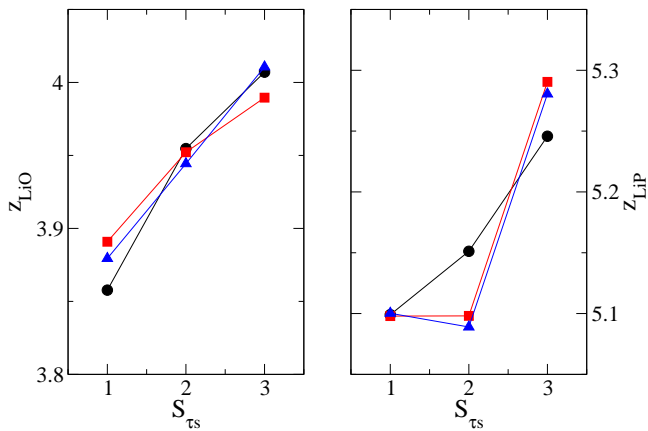}
\caption{Mean coordination numbers (a) $z_{\mathrm{LiO}}$ and (b)
$z_{\mathrm{LiP}}$ for lithium ions residing at sites with short
($S_{\tau s}\!=\!1$), medium ($S_{\tau s}\!=\!2$) and long ($S_{\tau
s}\!=\!3$) mean residence times $\tau_{s}$. Each subset
contains about 1/3 of all sites. For example, $S_{\tau s}\!=\!1$
consist of sites with residence times $\tau_s\!<\!\tau_1$, where
$C(\lg \tau_1)\!=\!1/3$.} \label{fig4}
\end{figure}

The relation between the lithium jump dynamics and the local glass
structure can be studied when we analyze the environments of sites
with diverse mean residence times. Therefore, we now compare the mean
coordination numbers of lithium ions at sites with short ($S_{\tau
s}\!=\!1$), medium ($S_{\tau s}\!=\!2$) and long ($S_{\tau s}\!=\!3$)
residence times $\tau_s$. In Fig.~\ref{fig4}, we see
that, on average, lithium ions at sites with short $\tau_s$ exhibit
smaller mean coordination numbers $z_{\mathrm{LiO}}$ and
$z_{\mathrm{LiP}}$ than those ions at sites with long $\tau_s$. This
suggests that sites featuring slow lithium jumps are well embedded in
the phosphate-glass matrix, however, the effects are weak. For the
mean coordination number $z_{\mathrm{LiLi}}$, we observe no
significant correlation. Also, we find no systematic effects due to
other quantities characterizing the local lithium environments.
Hence, our results give no evidence that any particular structural
property determines the lithium jump dynamics in LiPO$_3$ glass.
However, the local structure determines the local energy landscape.
We find that the average energy of a lithium ion at a site, or,
equivalently, the site energy $\varepsilon_s$ is relatively low for
sites where the lithium ions have oxygen neighbors that contain a
high fraction of NBO. In the following, we directly study the effects
of the local energy landscape on lithium jump dynamics.

\begin{figure}
\includegraphics[angle=0,width=8.5cm,clip]{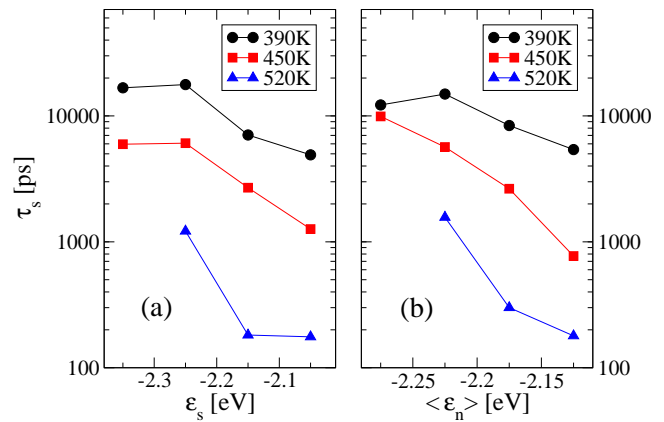}
\caption{Mean residence times $\tau_s$ at the lithium sites plotted
as a function of (a) the site energy $\varepsilon_s$ and (b) the
average site energy of the neighboring sites,
$\langle\varepsilon_n\rangle$. }\label{fig5}
\end{figure}

To determine the site energy $\varepsilon_s$ of a given site and for
a given $T$, we identify the lithium ion, if any, that occupies this
site in a considered MD configuration at this $T$, calculate its
potential energy using Eq.~\ref{potential} and average over many
configurations. Figure \ref{fig5}(a) shows $\tau_s$ as a function of
the site energies $\varepsilon_s$. As may be expected, the mean
residence time decreases with increasing $\varepsilon_s$. However,
decays by factors of between five and ten are small compared to the
width of $G(\lg \tau_s)$, which spans at least three orders of
magnitude, see Fig.~\ref{fig3}. For example, if a random trap model
were an appropriate description, i.e., if the residence time were
exclusively determined by $\varepsilon_s$, a decline of $\tau_s$ by a
factor of $\sim 2000$ would be expected based on $\Delta
\tau_s\!=\!\exp[\Delta \varepsilon_s/(k_BT)]$, $\Delta
\varepsilon_s\!=\!0.3\mathrm{\,eV}$ and $T\!=\!450\mathrm{\,K}$.
Thus, the jump rate at a site depends not only on $\varepsilon_s$,
but there are additional factors. For example, the energy barriers
separating the sites may show a distribution of barrier heights, as
expected for the energy landscape of a disordered material.

Another important factor can be inferred from an inspection of
Fig.~\ref{fig5}(b), where the mean residence time at a site is
plotted as a function of the average site energy,
$\langle\varepsilon_n\rangle$, of all \emph{neighboring} sites. Here,
we define neighboring sites based on the criterion that their
center-of mass distance $r_{cm}$ be smaller than the distance
corresponding to the first minimum of the Li-Li pair correlation
function. Interestingly, we observe that $\tau_s$ strongly decreases
with increasing $\langle\varepsilon_n\rangle$, indicating that, on
average, sites surrounded by high energetic neighbors feature fast
lithium dynamics. On first glance, this result is counterintuitive
since one would expect that jumps to sites with comparatively high
site energies are characterized by small rates and, hence, sites with
high values of $\langle\varepsilon_n\rangle$ should exhibit long
$\tau_s$.

\begin{figure}
\includegraphics[angle=0,width=8.5cm,clip]{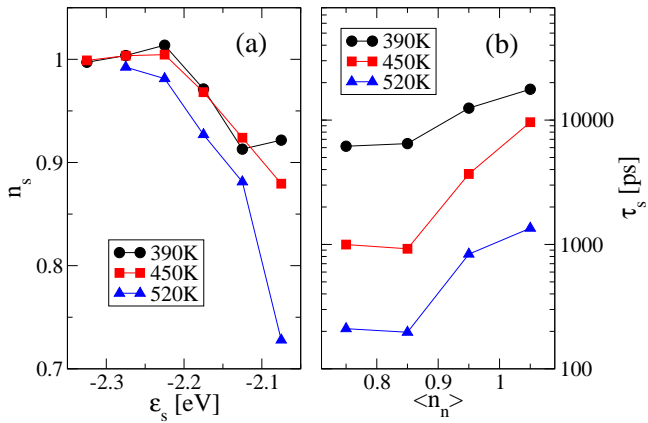}
\caption{(a) Mean number $n_s$ of lithium ions at a site as a
function of the site energy $\varepsilon_s$. (b) Mean residence time
$\tau_s$ at a lithium site as a function of $n_s$.}\label{fig6}
\end{figure}

However, a decrease of $\tau_s$ with $\langle\varepsilon_n\rangle$
can be understood when we consider our result that a vacancy
mechanism dominates the redistribution of a small fraction of empty
sites. In such scenario, the ionic jump rates are first of all
limited by the availability of a vacant neighboring site. In
Fig.~\ref{fig6}(a), we see that the occupation numbers $n_s$ are
smaller for sites with high site energies $\varepsilon_s$, as may be
expected. Thus, when the neighbors of a site A exhibit high
site energies, these neighbors are often unoccupied so that ions at
site A frequently have the chance for a jump.
Consequently, in our case of a vacancy mechanism, it is plausible
that sites characterized by high values of
$\langle\varepsilon_n\rangle$ show relatively short $\tau_s$, as is
observed in Fig.~\ref{fig5}(b). To further check our argumentation,
we plot the mean residence time at a site as a function of the
average occupation number, $\langle n_n\rangle$, of all neighboring
sites in Fig.~\ref{fig6}(b). We see that $\tau_s$ strongly grows with
$\langle n_n\rangle$, supporting our conclusion that lithium
diffusion in LiPO$_3$ glass is governed by the competition of the
ions for a small fraction of vacant sites. These results clearly show
that single-particle approaches do not allow a complete understanding
of ion dynamics in glasses.

\subsection{Correlated back-and-forth jumps}

The results of our previous work~\cite{Vogel} and the discussion of
Fig.~\ref{fig3} imply that correlated back-and-forth jumps are
another feature of lithium dynamics in the studied model. To
investigate this effect in more detail, we calculate the backjump
probability $P^{b}$ of finding a lithium ion at the same site after
exactly two jumps. Analyzing all jumps during our production runs, we
find that the backjump probability increases from $P^{b}\!=\!0.77$ at
$T\!=\!520\mathrm{\,K}$ to $P^{b}\!=\!0.91$ at
$T\!=\!390\mathrm{\,K}$. For comparison, a value
$P^{b}\!\approx\!0.25$ would result for a random walk since, on
average, each lithium site has about four neighboring sites. Thus,
the observed backjump probabilities clearly indicate that correlated
back-and-forth jumps are an important phenomenon in the studied
temperature range.

\begin{figure}
\includegraphics[angle=0,width=7.5cm,clip]{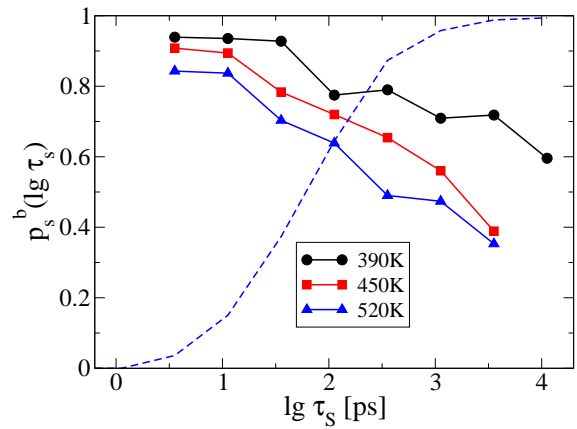}
\caption{Backjump probability $p_s^{b}$ as a function of the
logarithm of the mean residence time, $\lg \tau_s$, for various
temperatures. Dashed line: Cumulative distribution of the mean
residence time, $C(\lg\tau_s)$, for
$T\!=\!520\mathrm{\,K}$.}\label{fig7}
\end{figure}

To obtain further insights, we next determine the site-dependent
backjump probability $p_s^{\,b}$, i.e., for each site, we calculate
the probability that a jump from this site to another site is
followed by a direct backjump. Motivated by results of Heuer and
coworkers,~\cite{Lammert} we then study the relation between the
backjump probability $p_s^{\,b}$ and the mean residence time
$\tau_s$. From the results in Fig.~\ref{fig7}, it is evident that the
backjump probability is higher at lower $T$. Furthermore, we see that
$p_s^{\,b}$ strongly decreases with increasing $\lg \tau_s$.
Specifically, comparison with the cumulative distribution $C(\lg
\tau_s)$ shows that about 90\% of the jumps starting from sites with
short $\lg \tau_s$ are followed by a direct backjump, whereas
backjumps occur with nearly statistical probability,
$p_s^{\,b}\!\approx\!0.25$, after the escape from sites with long
$\tau_s$. Reinspecting also $F_q(\lg t)$ in Fig.~\ref{fig3}, these
findings show that lithium hopping on the time scale of the lithium
relaxation resembles a random walk, whereas ions that exit their
sites at much earlier times very often return to these sites, i.e.,
such jumps are unsuccessful. A similar behavior was observed for
lithium silicate glasses by Heuer and coworkers.~\cite{Lammert}

\begin{figure}
\includegraphics[angle=0,width=8.5cm,clip]{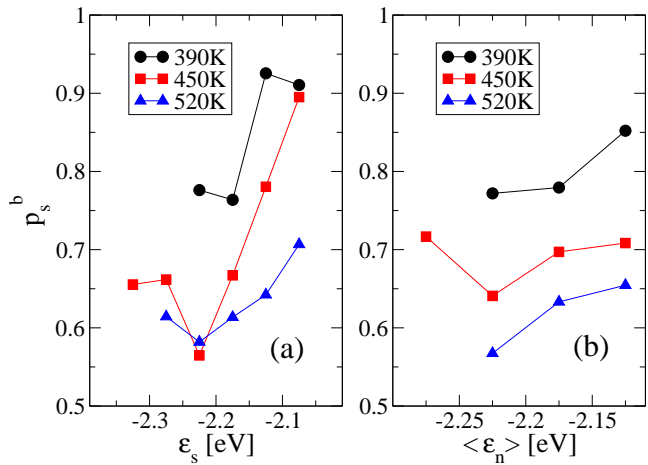}
\caption{Backjump probability $p_s^{b}$ as a function of (a) the site
energy $\varepsilon_s$ and (b) the average site energy of the
neighboring sites, $\langle\varepsilon_n\rangle$.}\label{fig8}
\end{figure}

Finally, we analyze the relation between the backjump probability and
the local energy landscape. In Fig.~\ref{fig8}(a), we see that, on
average, sites with high site energies $\varepsilon_s$ show somewhat
higher values of $p_s^{\,b}$. Figure~\ref{fig8}(b) displays
$p_s^{\,b}$ as a function of the average energy,
$\langle\varepsilon_n\rangle$, of the neighboring sites. Obviously,
the backjump probability hardly depends on
$\langle\varepsilon_n\rangle$. When we assume single-particle
motion in an energy landscape, these results are difficult to
understand. In such approach, jumps to sites with lower site energies
are favored and, hence, the backjump probability should be high for
sites characterized by small values of $\varepsilon_s$ and high
values of $\langle\varepsilon_n\rangle$, at variance with our
observations. Hence, our findings for the back-and-forth dynamics
again suggest that multiparticle interactions are important for the
lithium jumps in LiPO$_3$ glass. We add that we find no evidence for
a systematic dependence of $p_s^{\,b}$ on properties of the local
glass structure such as the mean coordination numbers
$z_{\mathrm{LiO}}$, $z_{\mathrm{LiP}}$ and $z_{\mathrm{LiLi}}$.

\section{Discussion and Summary}

We performed MD simulations to investigate lithium dynamics in
LiPO$_3$ glass at temperatures where the phosphate matrix is
basically rigid on the time scale of our simulation. We observed that
lithium ionic migration in this glassy matrix results from jumps
between well defined sites. Making use of a straightforward
algorithm,~\cite{Lammert} the lithium sites were completely
identified from the ionic trajectories. The results show that the
sites are typically compact objects of similar size and globular
shape. Moreover, the number of lithium sites hardly outnumbers the
number of lithium ions, as was observed for lithium silicate
glasses.~\cite{Lammert} Characterizing the structural and dynamical
features of the sites, we found that these properties are unmodified
on the time scale of the lithium ionic relaxation. This allowed us to
systematically study the relationship between structure and dynamics.

Analyzing the phosphate dynamics in detail, we found that the glassy
matrix, apart from temporary fluctuations, is static on the
$20\,\mathrm{ns}$-time scale of our simulation. In particular, there
is a basically time independent spatial distribution of NBO. While
one may speculate whether the temporary matrix fluctuations assist
the lithium jumps, as was observed for other ionic
glasses,~\cite{Sunyer3,Angell} the present results are at variance
with a substantial structural relaxation of the glassy matrix, e.g.,
a redistribution of NBO, as was proposed in previous
work.~\cite{Greaves1,Maass3} Instead, the properties of the lithium
sites are determined by the configuration frozen in at the glass
transition. In particular, we observed that the lithium sites are
located in regions with high local concentrations of NBO so that most
lithium ions at a site are exclusively surrounded by this oxygen
species. Though we do not expect that our findings are unique to
LiPO$_3$ glass, we cannot exclude that the scenario changes for
small alkali concentrations, where NBO may be formed along the alkali
trajectories.~\cite{Oviedo}

Mapping the lithium trajectories onto sequences of jumps between the
identified sites, we analyzed the mechanism for the ionic diffusion.
The mean jump times and the mean jump lengths indicated that lithium
migration results from hopping between neighboring sites. A
statistical analysis of the repopulation of the lithium sites showed
that the sites are usually first vacated by one ion and shortly
afterwards entered by another ion, implying that lithium hopping is
governed by a vacancy mechanism. Such mechanism is reasonable in a
situation where the number of alkali sites is only slightly bigger
than the the number of alkali ions, as is the case both in lithium
silicate glasses~\cite{Lammert} and in our model of a lithium
phosphate glass. For silicate glasses, a vacancy-like mechanism was
also observed when inspecting examples of ionic
trajectories.~\cite{Cormack}

To characterize the average dynamical behavior of a lithium ion at a
particular site, we determined the mean residence times $\tau_s$ at
the sites. We found that the sites feature a broad distribution
$G(\lg \tau_s)$, indicating the existence of pronounced dynamic
heterogeneities. $G(\lg \tau_s)$ broadens upon cooling 
so that the distribution extends over more
than four orders of magnitude at the lowest $T$ studied.
Consistently, we observed in our previous MD study of LiPO$_3$
glass~\cite{Vogel} that dynamic heterogeneities increasingly
contribute to the nonexponential lithium relaxation upon cooling. All
these simulation results are in agreement with findings from
multidimensional NMR experiments on silver phosphate
glasses,~\cite{Vogel_JNCS,Vogel_PCCP} which show that a very broad
distribution of jump rates governs silver dynamics at low
$T$, where the silver relaxation occurs on the time scale of
$\mathrm{ms\!-\!s}$.

The present approach provides direct access to the origin of the
dynamic heterogeneities. We observed that, on average, lithium ions
at sites with long $\tau_s$ show somewhat higher coordination numbers
$z_{\mathrm{LiO}}$ and $z_{\mathrm{LiP}}$, suggesting that these
sites are located in a matrix-rich environment. However, the direct
effects of the local glass structure on the lithium jump dynamics are
weak. On the other hand, the local structure determines the local
energy landscape, which in turn strongly affects the residence times.
Specifically, as may be expected, we found that $\tau_s$ decreases
with the site energy $\varepsilon_s$. More interestingly, $\tau_s$
also declines when the average site energy
$\langle\varepsilon_n\rangle$ of the neighboring sites increases.
Such behavior is difficult to capture within single-particle
approaches on ion transport in glasses, including percolation
approaches. However, it can be understood when we consider our
results that there is a small fraction of empty sites that are
redistributed in a vacancy mechanism. In such scenario, the jump
rates are limited by the restricted access to a vacant site. Due to a
correlation between the energies $\varepsilon_s$ and the occupation
numbers $n_s$ of the sites, the probability of finding a vacant
neighboring site is high when the latter exhibit high site energies
and, consequently, $\tau_s$ is expected to decrease with
$\langle\varepsilon_n\rangle$, in agreement with our observation. We
conclude that the pronounced dynamic heterogeneities in LiPO$_3$
glass result from both a disordered energy landscape and a
distribution of probabilities of having access to a vacant
neighboring site.

We also calculated the site-dependent backjump probability
$p_s^{\,b}$ that a jump from a particular site to another site is
followed by a direct backjump. Such analysis allowed us to quantify
the importance of correlated back-and-forth jumps. We found that
$p_s^{\,b}$ strongly depends on the mean residence time $\tau_s$,
i.e., on the dynamical state of an ion. High values of $p_s^{\,b}$
for sites with short $\tau_s$ indicate that correlated back-and-forth
jumps are a relevant feature of lithium dynamics at the studied $T$.
Specifically, backjump probabilities $p_s^{\,b}\!\approx\!0.9$ that
increase upon cooling show that only $\sim10\%$ of the jumps starting from
sites with short $\tau_s$ are successful and, hence, ion transport is
substantially slowed down due to back-and-forth jumps, in particular
at low $T$. In contrast, sites characterized by long $\tau_s$ show
backjump probabilities that are close to the value expected for a
random jump. These results for LiPO$_3$ glass are in agreement with
findings for lithium silicate glasses by Heuer and
coworkers,~\cite{Lammert} implying that there is a similar mechanism
for lithium jump dynamics in different types of glasses. However,
correlated back-and-forth jumps were not observed in a computational
study of ion dynamics at comparatively high $T$.~\cite{Kob} This
apparent discrepancy may be reconciled by our result that the
importance of this phenomenon decreases with increasing $T$.

Our results may also help to unravel two puzzling effects related to
ion dynamics. As aforementioned, applications of different
experimental techniques lead to different conclusions about the
relevance of correlated back-and-forth jumps. These discrepancies may
be explained by the strong dependence of the backjump probability on
the dynamical state of an ion. Similar to Heuer and
coworkers,~\cite{Lammert} we argue that the observed correlated
back-and-forth jumps of the fast ions dominate the dispersive regime
of the electric conductivity
$\sigma(\nu)$.~\cite{Funke2,Jonscher,Martin,Funke3,Roling} An
additional contribution to the dispersion may result from fast
forward-backward motions on length scales shorter than the Li-Li
interatomic distance, e.g., jumps between sites and their satellites,
which are not considered in the present analysis. In contrast,
studying the dynamical behavior of subensembles of slow ions in 
multidimensional NMR
experiments,~\cite{Vogel_PCCP,Vogel_PRB} no evidence for correlated
back-and-forth jumps was found. Consistently, we observed that
lithium dynamics on the time scale of the lithium relaxation
resembles a random walk.

The second puzzling effect concerns the approximate time-temperature
superposition observed for the incoherent intermediate scattering
function $F_q(t)$ in simulation studies.~\cite{Heuer1,Vogel} This
behavior is surprising since both dynamic
heterogeneities~\cite{Heuer2,Vogel} and back-and-forth jumps become
more important with decreasing $T$, see Figs.~\ref{fig3} and
\ref{fig7}, and, hence, one might be tempted to conclude that the
nonexponentiality of $F_q(t)$ should increase upon cooling. However,
such simple argumentation is only possible when the backjump
probability is independent of the dynamical state. Here, the
back-and-forth jumps of the fast ions shift the time $t_0$ when
$F_q(t)$ starts to decay to longer times, while the time $t_1$ when
the decay of $F_q(t)$ is complete is not affected due to
back-and-forth dynamics because no enhanced back-jump probability
exists for the slow ions. This means that, in our case,
back-and-forth jumps \emph{reduce} the time interval $t_1\!-\!t_0$,
i.e., the nonexponentiality of $F_q(t)$. Hence, when $T$
is lowered, there are two competing effects. While the increasing
heterogeneity of dynamics leads to a stronger nonexponentiality, the
growing backjump probability of the fast ions delays the onset of the
decay, $t_0$, further and further and, hence, reduces the
nonexponentiality. Thus, both effects may compensate each other,
resulting in a nearly temperature independent nonexponentiality of
$F_q(t)$.

In summary, we found that, in a model of LiPO$_3$ glass, the disordered 
energy landscape provided by the phosphate-glass matrix plays a
major role for the lithium jump dynamics. For a complete understanding 
of the jump-diffusion mechanism, single-particle approaches are not
sufficient, but rather it is importont to consider the competition of 
the lithium ions for a small fraction of vacant sites at every instant. 
As a consequence, the dynamical behavior is highly complex. In 
particular, pronounced dynamic heterogeneities exist and the 
backjump probability depends on the dynamical state of an ion.

\begin{acknowledgments}
I thank S.\ C.\ Glotzer for a generous grant of computer time and A.\
Heuer, H.\ Lammert and J.\ Reinisch for many stimulating discussions.
Funding of the Deutsche Forschungsgemeinschaft through the Emmy
Noether-Programm is gratefully acknowledged.
\end{acknowledgments}


\begin{thebibliography}{99}
\bibitem{BIM} A.\ Bunde, M.\ D.\ Ingram and P.\ Maass, J.\ Non-Cryst.\ Solids 172-174, 1222 (1994)
\bibitem{Ngai} K.\ L.\ Ngai, J.\ Non-Cryst.\ Solids 203, 232 (1996)
\bibitem{BFI} A.\ Bunde, K.\ Funke and M.\ D.\ Ingram, Solid State Ionics 105, 1 (1998)
\bibitem{Funke1} K.\ Funke, Prog.\ Solid State Chem.\ 22, 111 (1993)
\bibitem{Elliott} S.\ R.\ Elliott, Solid State Ionics 70/71, 27 (1994)
\bibitem{Funke2} K.\ Funke, B.\ Roling and M.\ Lange, Solid State Ionics 105, 195 (1998)
\bibitem{Stevels} J.\ M.\ Stevels, in \emph{Handbuch der Physik}, Vol.\ 20, edited by S.\ Fl\"ugge, Springer Verlag, Berlin (1957)
\bibitem{Ingram} M.\ D.\ Ingram, Philos.\ Mag.\ B 60, 729 (1989)
\bibitem{Dyre1} J.\ C.\ Dyre and T.\ B.\ Schr\o der, Rev.\ Mod.\ Phys.\ 72, 873 (2000)
\bibitem{Svare} I.\ Svare, F.\ Borsa, D.\ R.\ Torgeson and S.\ W.\ Martin, Phys.\ Rev.\ B 48, 9336 (1993)
\bibitem{Sen} S.\ Sen, A.\ M.\ George and J.\ F.\ Stebbins J.\ Non-Cryst.\ Solids 197, 53 (1996)
\bibitem{Baranovskii} S.\ D.\ Baranovskii and H.\ Cordes, J.\ Chem.\ Phys.\ 111, 7546 (1999)
\bibitem{Maass1} P.\ Maass, J.\ Petersen, A.\ Bunde, W.\ Dieterich and H.\ E.\ Roman, Phys.\ Rev.\ Lett.\ 66, 52 (1991)
\bibitem{Maass2} P.\ Maass, M.\ Meyer, A.\ Bunde and W.\ Dieterich, Phys.\ Rev.\ Lett.\ 77, 1528 (1996)
\bibitem{Dietrich} D.\ Kn\"odler, P.\ Penzig and W.\ Dieterich, Solid State Ionics 86-88, 29 (1996)
\bibitem{Greaves1} G.\ N.\ Greaves and K.\ L.\ Ngai, Phys.\ Rev.\ B 52, 6358 (1995)
\bibitem{Greaves2} G.\ N.\ Greaves, J.\ Non-Cryst.\ Solids 71, 203 (1985)
\bibitem{Maass3} P.\ Maass, A.\ Bunde and M.\ D.\ Ingram, Phys.\ Rev.\ Lett.\ 68, 3064 (1992)
\bibitem{Moynihan} C.\ T.\ Moynihan, L.\ P.\ Boesch and N.\ L.\ Laberge, Phys.\ Chem.\ Glasses 14, 122 (1973)
\bibitem{Liu} C.\ Liu and C.\ A.\ Angell, J.\ Non-Cryst.\ Solids 83, 162 (1986)
\bibitem{Green1} P.\ F.\ Green, D.\ L.\ Sidebottom and R.\ K.\ Brow, J.\ Non-Cryst.\ Solids 172-174, 1353 (1994)
\bibitem{Green2} P.\ F.\ Green, E.\ F.\ Brown and R.\ K.\ Brow, J.\ Non-Cryst.\ Solids 255, 87 (1999)
\bibitem{12} R.\ B\"ohmer, R.\ V.\ Chamberlin, G.\ Diezemann, B.\ Geil, A.\ Heuer, G.\ Hinze, S.\ C.\ Kuebler, R.\ Richert, B.\ Schiener, H.\ Sillescu, H.\ W.\ Spiess, U.\ Tracht and M.\ Wilhelm, J.\ Non-Cryst.\ Solids 235-237, 1 (1998)
\bibitem{Jonscher} A.\ K.\ Jonscher, Nature 267, 673 (1977)
\bibitem{Martin} S.\ W.\ Martin and C.\ A.\ Angell, J.\ Non-Cryst.\ Solids 83, 185 (1986)
\bibitem{Funke3} K.\ Funke and C.\ Cramer, Curr.\ Opin.\ Solid State Mater.\ Sci.\ 2, 483 (1997)
\bibitem{Roling} B.\ Roling, A.\ Happe, K.\ Funke and M.\ D.\ Ingram, Phys. Rev.\ Lett.\ 78, 2160 (1997)
\bibitem{Bohmer1} R.\ B\"ohmer, T.\ J\"org, F.\ Qi and A.\ Titze, Chem.\ Phys.\ Lett.\ 316, 419 (2000)
\bibitem{Bohmer2} F.\ Qi, T.\ J\"org and R.\ B\"ohmer, Solid State Nucl.\ Magn.\ Reson.\ 22, 484 (2002)
\bibitem{Vogel_JNCS} M.\ Vogel, C.\ Brinkmann, H.\ Eckert and A.\ Heuer, J.\ Non-Cryst.\ Solids 307-310, 971 (2002)
\bibitem{Vogel_PCCP} M.\ Vogel, C.\ Brinkmann, H.\ Eckert and A.\ Heuer, Phys.\ Chem.\ Chem.\ Phys.\ 4, 3237 (2002)
\bibitem{Vogel_PRB} M.\ Vogel, C.\ Brinkmann, H.\ Eckert and A.\ Heuer, Phys.\ Rev.\ B (in press)
\bibitem{Smith} W.\ Smith, G.\ N.\ Greaves and M.\ J.\ Gillan, J.\ Chem.\ Phys.\ 103, 3091 (1995)
\bibitem{Bala} S.\ Balasubramanian and K.\ J.\ Rao, J.\ Non-Cryst.\ Solids 181, 157 (1995)
\bibitem{Karth} A.\ Karthikeyan and K.\ J.\ Rao, J.\ Phys.\ Chem.\ B 101, 3105 (1997)
\bibitem{Timpel} D.\ Timpel and K.\ Scheerschmidt, J.\ Non-Cryst.\ Solids 232-234, 245 (1998)
\bibitem{Kieffer} J.\ Kieffer, J.\ Non-Cryst.\ Solids 255, 56 (1999)
\bibitem{Park} B.\ Park and A.\ N.\ Cormack, J.\ Non-Cryst.\ Solids 255, 112 (1999)
\bibitem{Horbach1} J.\ Horbach, W.\ Kob and K.\ Binder, Chem.\ Geology 174, 87 (2001)
\bibitem{Habasaki} J.\ Habasaki and Y.\ Hiwatari, Phys.\ Rev.\ E 65, 021604 (2002)
\bibitem{Cormack} A.\ N.\ Cormack, J.\ Du and T.\ R.\ Zeitler, Phys.\ Chem.\ Chem.\ Phys.\ 4, 3193 (2002)
\bibitem{Heuer1} A.\ Heuer, M.\ Kunow, M.\ Vogel and R.\ D.\ Banhatti, Phys.\ Chem.\ Chem.\ Phys.\ 4, 3185 (2002)
\bibitem{Heuer2} A.\ Heuer, M.\ Kunow, M.\ Vogel and R.\ D.\ Banhatti, Phys.\ Rev. B.\ 66, 224201 (2002)
\bibitem{Vogel} M.\ Vogel, Phys.\ Rev.\ B 68, 184301 (2003)
\bibitem{Jund} P.\ Jund, W.\ Kob and R.\ Jullien, Phys.\ Rev.\ B 64, 134303 (2001)
\bibitem{Sunyer1} E.\ Sunyer, P.\ Jund and R.\ Jullien, Phys.\ Rev.\ B 65, 214203 (2002)
\bibitem{Kob} E.\ Sunyer, P.\ Jund, W.\ Kob and R.\ Jullien, J.\ Non-Cryst.\ Solids, 307-310, 939 (2002)
\bibitem{Horbach2} J.\ Horbach, W.\ Kob and K.\ Binder, Phys.\ Rev.\ Lett.\ 88, 125502 (2002)
\bibitem{Sunyer2} E.\ Sunyer, P.\ Jund and R.\ Jullien, J.\ Phys.: Condens.\ Matt.\ 15, S1659 (2003)
\bibitem{Vessal} B.\ Vessal, G.\ N.\ Greaves, P.\ T.\ Martin, A.\ V.\ Chadwick, R.\ Mole and S.\ Houde-Walter, Nature (London) 356, 504 (1992)
\bibitem{Lammert} H.\ Lammert, M.\ Kunow and A.\ Heuer, Phys.\ Rev.\ Lett.\ 90, 215901 (2003)
\bibitem{Karthikeyan} A.\ Karthikeyan, P.\ Vinatier, A.\ Levasseur and K.\ J.\ Rao, J.\ Phys.\ Chem.\ B 103, 6185 (1999)
\bibitem{Alam} J.-J.\ Liang, R.\ T.\ Cygan and T.\ M.\ Alam, J.\ Non-Cryst.\ Solids 263-264, 167 (2000)
\bibitem{Hoppe} U.\ Hoppe, G.\ Walter, R.\ Kranold and D.\ Stachel, J.\ Non-Cryst.\ Solids 263-264, 29 (2000)
\bibitem{Brow} R.\ K.\ Brow, J.\ Non-Cryst.\ Solids 263-264, 1 (2000)
\bibitem{Wullen} L.\ van W\"ullen, H.\ Eckert and G.\ Schwering, Chem.\ Mater.\ 12, 1840 (2000)
\bibitem{Muru} K.\ Muruganandam, M.\ Seshasayee and S.\ Patanaik, Solid State Ionics 89, 313 (1996)
\bibitem{English} S.\ English and W.\ E.\ S.\ Turner, J.\ Am.\ Ceram.\ Soc.\ 13, 182 (1930)
\bibitem{Dyre2} J.\ Dyre, J.\ Non-Cryst.\ Solids 307, 939 (2002)
\bibitem{Sunyer3} E.\ Sunyer, P.\ Jund and R.\ Jullien, J.\ Phys.: Condens.\ Matt.\ 15, L431 (2003)
\bibitem{Angell} C.\ A.\ Angell, L.\ Boehm, P.\ A.\ Cheeseman and S.\ Tamaddon, Solid State Ionics 5, 659 (1981)
\bibitem{Oviedo} J.\ Oviedo and J.\ F.\ Sanz, Phys.\ Rev.\ B 58, 9047 (1998)
\end{thebibliography}
\end{document}